\documentclass[]{spie}

\usepackage[utf8]{inputenc} 
\usepackage[T1]{fontenc}    
\usepackage{hyperref}       
\usepackage{url}            
\usepackage{booktabs}       
\usepackage{amsfonts}       
\usepackage{nicefrac}       
\usepackage{microtype}      
\usepackage{xcolor}         
\usepackage{graphics}
\usepackage{breqn}

\title{Synthesis of Annotated Colon Cancer Tissue Images from Gland Layout}

\author{Srijay Deshpande}
\author{Fayyaz Minhas}
\author{Nasir Rajpoot}
\affil{Tissue Image Analytics Centre, University Of Warwick, Coventry, United Kingdom}

\authorinfo{Further author information: (Send correspondence to Srijay Deshpande)\\ Srijay Deshpande: E-mail: deshpandesrijay@gmail.com}

\begin{document}

\maketitle

\section{Abstract}

Generating realistic tissue images with annotations is a challenging task that is important in many computational histopathology applications. Synthetically generated images and annotations are valuable for training and evaluating algorithms in this domain. To address this, we propose an interactive framework generating pairs of realistic colorectal cancer histology images with corresponding glandular masks from glandular structure layouts. The framework accurately captures vital features like stroma, goblet cells, and glandular lumen. Users can control gland appearance by adjusting parameters such as the number of glands, their locations, and sizes. The generated images exhibit good Frechet Inception Distance (FID) scores compared to the state-of-the-art image-to-image translation model. Additionally, we demonstrate the utility of our synthetic annotations for evaluating gland segmentation algorithms. Furthermore, we present a methodology for constructing glandular masks using advanced deep generative models, such as latent diffusion models. These masks enable tissue image generation through a residual encoder-decoder network.


\section{Introduction}

Deep learning algorithms necessitate large amounts of training data, which can be challenging and expensive to obtain, and may require involvement from highly skilled pathologists. Numerous techniques have been proposed for generating synthetic tissue images of high quality \cite{quiros2019pathology, fmahmooddeepsegmentation, safron, synclay}. Synthetic histology images are useful for developing predictive models for downstream tasks \cite{robustlabelsynthesize, Levine2020SynthesisOD, safron}, as well as for educational purposes \cite{educationapplication} and clinical quality assurance \cite{qualityassurance}. Most of the aforementioned use cases of synthetic histology image generation require image synthesis from custom glandular layouts or user-defined tissue parameters, such as the number of glands and disease grade.

In pathology image analysis, due to tissue heterogeneity and variations in acquired tissue images in laboratories, the data annotation phase must often be repeated for different tissue types, such as glands, fat tissues, and blood vessels, to achieve optimal performance. For tasks like gland segmentation, manually generating component masks that highlight glandular portions can be a laborious and time-consuming task. This presents a significant obstacle to generating a large amount of annotated data for segmentation algorithms. Some researchers have investigated generating synthetic pathology images from tissue component masks \cite{tissuecgan, sashimi, robustlabelsynthesize}. These methods either assume that input component masks are already present or require the explicit construction of component masks by generating random shapes of respective tissue components like nuclei, which can be erroneous and unrealistic. Furthermore, crafting component masks for larger multi-cellular structures, such as glands, can be challenging. Therefore, generating synthetic images along with component masks simultaneously is desirable, as it potentially reduces the cost of annotations and also produces realistic annotated pairs.

\begin{figure}[hbt!]
\centering
\includegraphics[width=\textwidth]{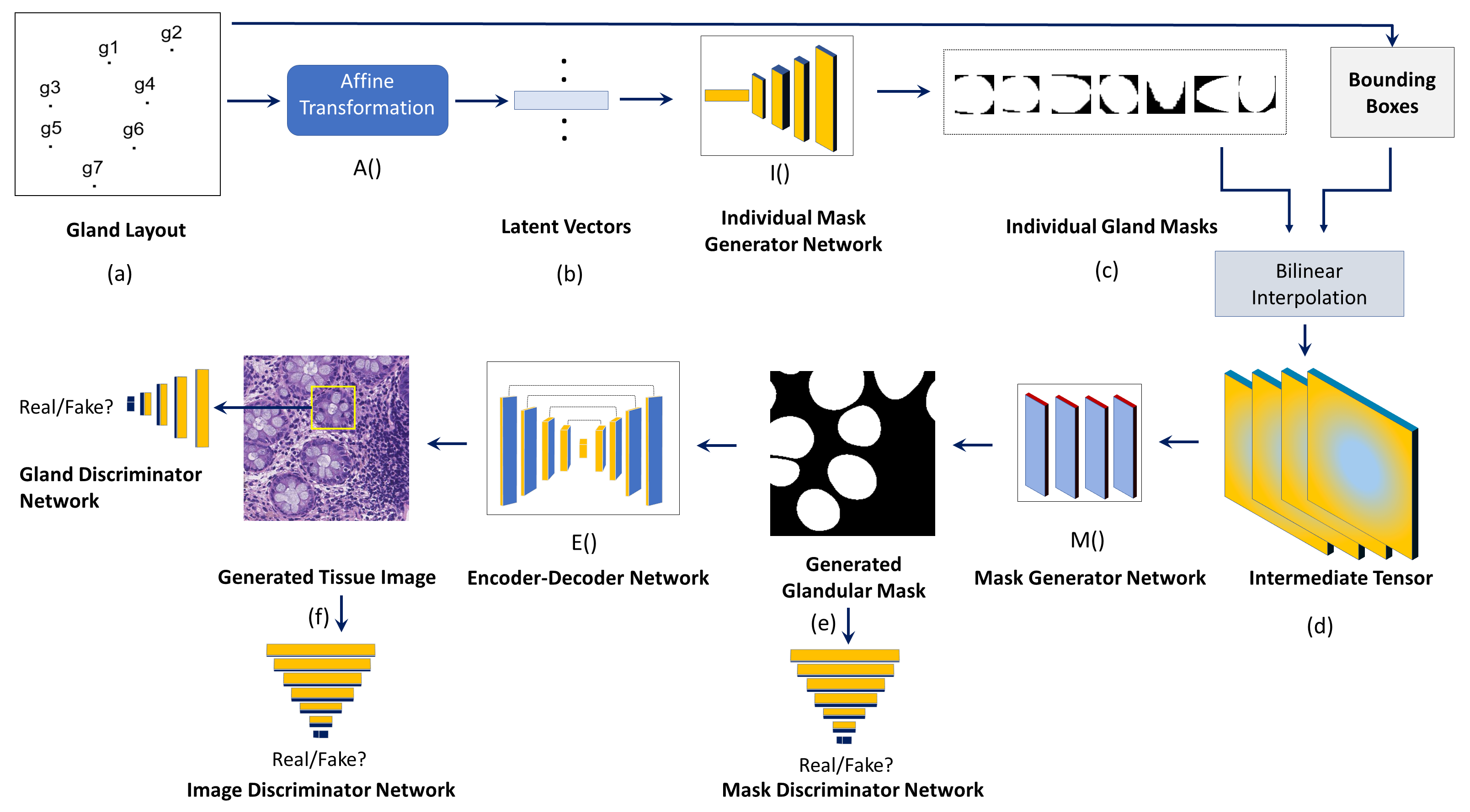}
\caption{Block diagram of the proposed framework: (a) shows an input gland layout where glands are arranged on 2-d spatial locations. Each gland is characterized by a gland specific vector which undergoes affine transformation to form latent vectors (b). Each latent vector is consumed by the individual mask generator which outputs binary individual glandular masks (c). The generated individual masks along with input bounding boxes act as an input to the bilinear interpolation algorithm, which wraps generated masks inside bounding boxes creating the intermediate tensor (d). The mask generator network consumes the intermediate tensor and generates the glandular mask (e), which is then passed through the encoder-decoder generator network generating the final tissue image (f). There are three discriminators employed for generated mask, image and glandular parts inside the image.}
\label{scene_generation_idea}
\end{figure}

In this work, we propose a user-interactive framework capable of generating colorectal tissue images along with corresponding tissue component masks simultaneously, using the input gland layout. To the best of our knowledge, it is the first framework that can generate annotated colorectal tissue images controlled from the gland layout. This layout enables users to specify the locations and sizes of the glands in colorectal tissue images. We harness the properties of Generative Adversarial Networks (GANs) within an adversarial setting to create realistic tissue images. Additionally, we present a methodology based on latent diffusion models \cite{latentdiffusion} for synthesizing glandular masks without relying on specific glandular layouts or similar inputs. Cutting-edge approaches, such as taming transformers \cite{tamingtransformers}, OpenAI's DALL-E \cite{dalle}, and Stable Diffusion \cite{latentdiffusion}, have also employed quantization-based architectures to produce lifelike images. Key highlights of our work are:

\begin{enumerate}

  \item We propose a framework that can generate realistic colorectal tissue images and corresponding tissue component/gland masks simultaneously. 
  
  \item The proposed framework allows user to change the appearance of glands by their locations and sizes. To the best of our knowledge, it is the first framework that can generate annotated colorectal tissue images controlled from the gland layout.
  
  \item We demonstrate the efficacy of the annotated pairs generated using the proposed framework for evaluation of the gland segmentation algorithm.

  \item We provide a vector quantized variational autoencoder (VQ-VAE) based methodology, supplemented by the integration of the Diffusion model. This combination enables the generation of realistic glandular masks, subsequently producing tissue images using the residual encoder decoder network.
  
\end{enumerate}

\section{Method}
\label{materialsmethods}

Our aim is to develop the framework to construct the colorectal cancer tissue image and its corresponding tissue component mask highlighting glandular regions, from the input gland layout. The gland layout can be described as a first quadrant Cartesian plane where users can arrange glands on its 2-d spatial locations. The gland layout is consumed by the framework $f$ that constructs the tissue component mask first, which later assists to generate the complete tissue image of size $N \times N$ pixels. The overview is given in Figure \ref{scene_generation_idea}. Below we describe the main components of the proposed framework:

The input gland layout can be assumed as a set of $n$ glands, $G_{layout} = \{g_k \equiv (\overrightarrow{l_k}, \overrightarrow{s_k}) \mid k = 1,2...n \}$, where $\overrightarrow{l_k}$ and $\overrightarrow{s_k}$ denote the location and size vectors respectively of the gland $g_k$. The gland $g_k$ is characterized by a gland specific vector $\overrightarrow{z_k}$ sampled from the Gaussian noise $z\sim \mathcal{N}(0,1)$. The Gaussian noise is used to ensure the variable appearance of glandular objects in the final image. The gland specific vector $\overrightarrow{z_k}$ is passed through the \textit{affine transformation} $A$, generating $n$ latent embeddings $\{ a_k \mid k = 1,2...n \}$ of dimensionality $D$,  i.e., $a_k = A(\overrightarrow{z_k};\theta_{A})$, where $\theta_{A}$ represents the function's trainable weights.

\subsection{Generation of Glandular Mask}
\label{sectioncomponentmask}

Generated latent embeddings are then consumed by the \textit{individual mask generator network} $I$, generating the corresponding individual gland binary masks $\{ m_k \mid k = 1,2...n \}$, each of the size $B \times B$ pixels i.e., $m_k = M(a_k;\theta_{I})$, where $\theta_{I}$ denotes the trainable parameters of the network. The \textit{individual mask generator network} is comprised of series of blocks having transpose convolution layer followed by the ReLU activation.

Next step is to align generated binary masks on the appropriate locations and construct the tissue component mask of same size as that of the final image, i.e. $N \times N$ pixels. For this purpose, we utilize the input bounding boxes for each object, $\{ b_k \mid k = 1,2...n \}$. These bounding boxes are either obtained from the datasets (procedure given in section \ref{experiments}) or realized from the input location and size parameters (section \ref{appendixc}). Each gland specific vector $\overrightarrow{z_k}$ is multiplied element-wise with the individual glandular mask $m_k$, and wrapped to the positions of bounding box $b_k$ using the fixed bilinear interpolation function\cite{bilinearinterpol} $F$, to give the intermediate tensor of dimensionality $D \times B \times B$ , i.e., $C = F(\{\overrightarrow{z_k},m_k,b_k \mid k = 1,2...n\})$. The intermediate tensor has $D$ channels which are then reduced to 1 channel using the \textit{mask generator network} $M$, to give the glandular mask or tissue component mask, i.e. $T = M(C,\theta_M)$, where $\theta_M$ denotes its trainable parameters.

\subsection{Tissue Image Generation}

After generating the tissue component mask, we feed it to the \textit{encoder-decoder generator network} $E$. The image-to-image translation encoder-decoder network is used as an image generator, to construct the final tissue image $Z = E(T,\theta_E)$. 

The encoder consists of a series of (convolution + ReLU) blocks that generate a fixed size encoding of the input image. The decoder constructs the final image using a series of (Transpose convolution + ReLU) blocks. Taking inspiration from Pix2Pix\cite{pix2pix} , the network $E$ also adopted skip-connections between the layers with the same sized feature maps so that the first downsampling layer is connected with the last upsampling layer, the second downsampling layer is connected with the second last upsampling layer, and so on. These skip-connections give image generator the flexibility to bypass the encoding part to subsequent layers and enable consideration of low level features from earlier encoding blocks in the generator.

\subsection{Discriminators}
\label{discriminators_section}

We employ 3 discriminator neural networks in an adversarial training setting: \textit{mask discriminator} $D_M(M,\theta_{M_D})$ for the generated glandular mask $(M)$, \textit{image discriminator} $D_Z(Z,\theta_{Z_D})$ for the generated tissue image $Z$, and the \textit{gland discriminator} $D_G(Z_{g_i},\theta_{G_D})$ for glandular portions $\{Z_{g_i} \mid i=1,2..n\}$ inside the tissue image, where $n$ is the number of glands; $\theta_{M_D}$, $\theta_{Z_D}$ and $\theta_{G_D}$ denotes the respective trainable parameters of those discriminators. The first two discriminators employ PatchGAN\cite{pix2pix} discriminator which predicts the realism of the different portions from the generated component mask and the tissue image, respectively. The adversarial losses based on these discriminators ensure tissue component masks and tissue images are realistic. 

The architecture of the \textit{gland discriminator} is comprised of a series of convolution operations and predicts a single score of realism for the generated glandular portions cropped out from the final tissue image based on input bounding boxes, and resized to a fixed size using bilinear interpolation\cite{bilinearinterpol}. It ensures the generated glands, one by one, appear real with their micro components like goblet cells and lumen.

We employ an adversarial loss function\cite{goodfellow2014generative} for all discriminators. A discriminator $D_t(X,\theta_{t_D})$ attempts to maximize the loss by classifying the input image $X$ generated by the generator function $G(X,\theta_G)$ which tries to minimize it, where $\theta_{t_D}$ and $\theta_G$ denotes the set of trainable parameters of the respective networks. $D_t$ is the discriminator type (of $D_M$, $D_Z$ or $D_G$). The adversarial min-max loss function is given by,

\begin{equation}
\min_{\theta_G} \max_{\theta_{t_D}} L_{GAN}^t(\theta_G,\theta_{t_D}) = E_{X\sim p_{data}(X)}[log D_{t}(X,\theta_{t_D})] + E_{X \sim p_{X}(X)}[log(1-D_{t}(G(X,\theta_G),\theta_{D_t})]
\label{advloss}
\end{equation}

The exact architectures of all of the above networks can be found in Appendix \ref{appendix}.

\subsection{Loss Components}
\label{loss_components}

The training loss of the proposed framework is made up of several terms as it involves multiple networks. The different loss components used in the framework
are described below:

The complete framework with trainable parameters $\{\theta_A, \theta_I, \theta_M, \theta_E, \theta_{M_D}, \theta_{Z_D}, \theta_{G_D} \}$ is trained by minimizing a loss function with the following components:\\

\noindent \textbf{Individual Binary Glandular Masks Reconstruction Loss:} This component penalize the difference between ground truth $\{ \hat{m_k} \mid k = 1,2...n \}$ and generated individual binary glandular masks $\{ m_k \mid k = 1,2...n \}$ using the mean square error (MSE) as follows,

  \begin{equation}
      L_{GlandMaskRec}(\theta_A, \theta_I) =  \sum_{k=1}^n MSE(\hat{m_k},m_k)
  \end{equation}
where $\hat{m_k}$ is the ground truth, $m_k$ is the generated individual glandular mask and $n$ is the number of glands in the tissue image. As we saw in section \ref{materialsmethods}, $m_k$ is dependent on the trainable parameters $\theta_A$ and $\theta_I$. Similarly, we can define the other loss components.\\

\noindent \textbf{Mask Reconstruction Loss:} This loss is employed to penalize the difference between ground truth $\hat{T}$ and generated tissue component mask $T$ with mean square error (MSE) as below,

  \begin{equation}
      L_{TissueMaskRec}(\theta_A, \theta_I, \theta_M) =  MSE(\hat{T},T)
  \end{equation}
where $\hat{T}$ is ground truth and $T$ is the generated tissue component mask. \\

\noindent \textbf{Image Reconstruction Loss:} This component captures the reconstruction error between ground truth $\hat{Z}$ and generated tissue image $Z$ using the $L1$ difference,

  \begin{equation}
      L_{ImageRec}(\theta_A, \theta_I, \theta_M, \theta_E) =  \| \hat{Z} - Z\|_1
  \end{equation}
where $\hat{Z}$ is ground truth and $Z$ is the generated tissue image. \\

\noindent \textbf{Adversarial Loss Components}: As we discussed in section \ref{discriminators_section}, we employ 3 adversarial loss components: $L_{GAN}^M$, $L_{GAN}^Z$ and $L_{GAN}^G$ for tissue component mask, tissue image and the glandular portions cropped out from the tissue image, respectively. Their expressions can be realized by putting $t=M$, $t=Z$ and $t=G$ in (\ref{advloss}). 

Thus, the overall learning problem can be cast as a the following adversarial optimization problem based on the linear combination of adversarial and reconstruction losses,

\begin{dmath}
\min_{\theta_A, \theta_I, \theta_M, \theta_E} \; \max_{\theta_{M_D},\theta_{Z_D},\theta_{G_D}} \lambda_1 L_{ImageRec}(\theta_A, \theta_I, \theta_M, \theta_E) + \lambda_2 L_{TissueMaskRec}(\theta_A, \theta_I, \theta_M) + \lambda_3 L_{GlandMaskRec}(\theta_A, \theta_I) + \lambda_4 L_{GAN}^M (\theta_G = \{\theta_A, \theta_I, \theta_M\}, \theta_{t_D} = \theta_{M_D}) + \lambda_5 L_{GAN}^Z (\theta_G = \{\theta_A, \theta_I, \theta_M, \theta_E\}, \theta_{t_D} = \theta_{Z_D}) + \lambda_6 L_{GAN}^G (\theta_G = \{\theta_A, \theta_I, \theta_M, \theta_E\}, \theta_{t_D} = \theta_{G_D})
\label{losseq}
\end{dmath}
where $\lambda_1, \lambda_2 .. \lambda_6$ denote the weights of corresponding loss components.

\subsection{Synthesis of Glandular Masks using Latent Diffusion Model}
\label{masksynthesis}

In this section, we discuss the latent diffusion model\cite{latentdiffusion} based framework used for synthesis of glandular masks. The process is done in two steps: (i) Training VQ-VAE \cite{vqvae} on glandular masks to learn the discrete latent representations, and (ii) sampling new latent representations using the Diffusion model \cite{diffusion1} conditioned on the cancer type (benign or malignant). Below we describe both steps in detail. 

\subsubsection{Generation of Discrete Latent Representations using VQ-VAE}

The VQ-VAE \cite{vqvae} model first passes the input glandular mask through an encoder neural network, which generates an encoded latent representation of the mask. The encoded latent representation is then passed through a quantization layer. This layer maps the continuous values of the latent representation to discrete values using a predefined codebook of vectors. Each vector in the codebook represents a possible value of the latent representation. The quantization layer assigns a vector from the codebook to each entity based on the Euclidean distance between the entity and the vectors in the codebook. The vectors obtained after the quantization layer are called quantized embeddings.

Finally, the quantized embeddings are passed through a decoder network, which reconstructs the glandular mask. In order to sample out-of-dataset masks, we acquire knowledge about the underlying latent variables. By learning a prior over these latents, we can significantly reduce the memory and computational resources required. This enables us to generate high-resolution masks simply by adjusting the size of the latent space.

\subsubsection{Sampling Glandular Masks using Diffusion Model}

To generate masks from a learnt distribution of latent variables or compressed masks, we establish a prior distribution over them. We achieve this using the Diffusion model\cite{diffusion1}. Diffusion models work in two phases. In the first phase known as forward diffusion, they use diffusion processes to transform real images into noisy images. These noisy images generally follow a Gaussian distribution. The second phase, reverse diffusion process, involves iteratively applying a denoising function or a deep neural network to the initial noise sampled from the Gaussian distribution, which removes noise and updates the image. This process is repeated multiple times to generate samples from the target distribution. Diffusion models use a learned denoising function, typically implemented as U-Net \cite{unet}, to denoise the image at each step of the reverse diffusion process. 

We apply the Diffusion model on latent vectors learnt using VQ-VAE \cite{vqvae} model, conditioned on the cancer type (benign or malignant). We enable this conditioning mechanism using the cross-attention \cite{latentdiffusion} mechanism. The sampled latent vectors are passed through the VQ-VAE decoder model trained in the last step to generate realistic glandular masks. In this experiment, instead of binary glandular masks, we use tissue component masks highlighting glands (in green color), stromal region (red color) and the background (blue color) as shown in Figure \ref{diffusion_images_annotated}. The idea is to model dependencies between maximum number of components inside the tissue image for crafting realistic glandular masks. The binary glandular masks can be obtained later from the generated tissue component masks by extracting glands and keeping others as the background. To generate the synthetic tissue images, we employ the residual encoder decoder network\cite{interactionscenegraph}. 

\begin{figure}[hbt!]
\centering
\includegraphics[width=380pt, keepaspectratio]{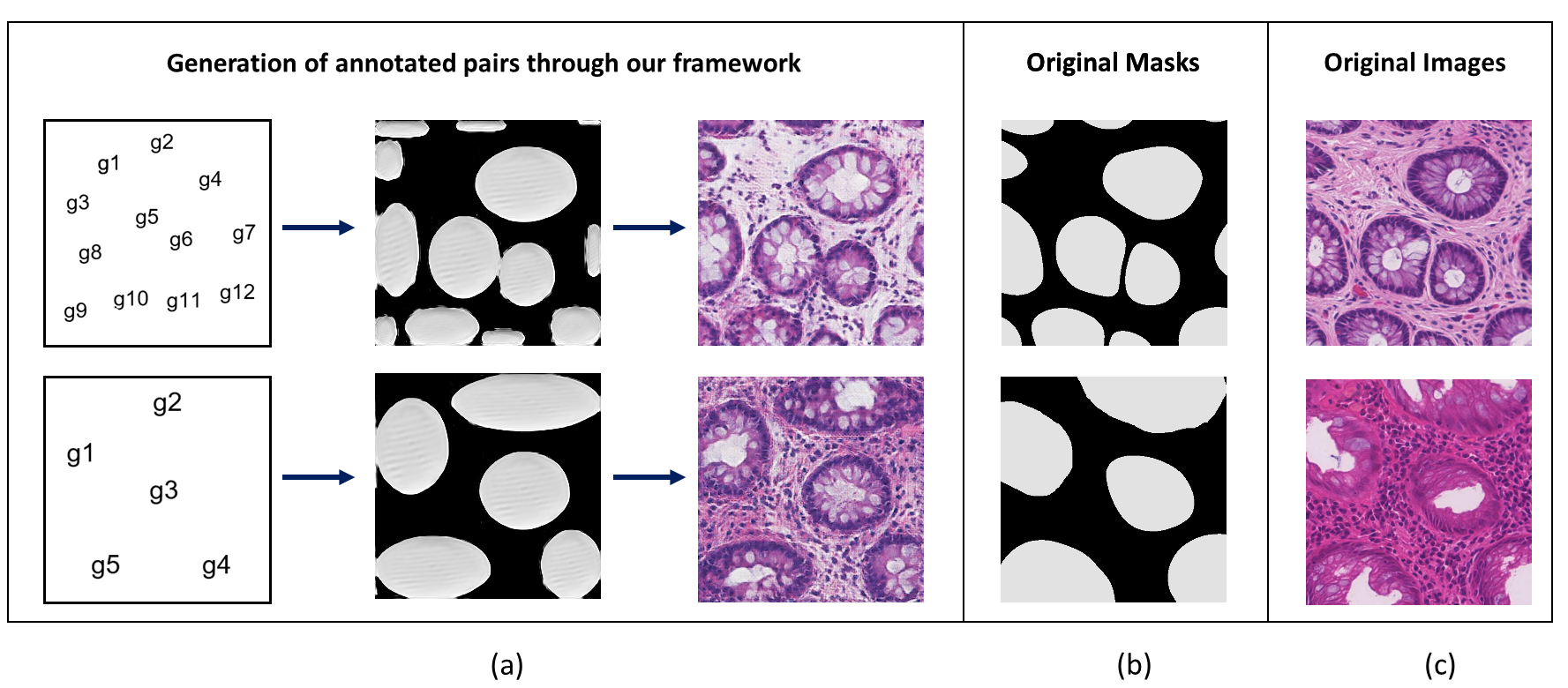}
\caption{Visual results of generated colorectal tissue images along with their gland segmentation masks from input gland layouts (a). (b) shows original gland segmentation masks while (c) shows the ground truth tissue images.}
\label{benign_results}
\end{figure}

\section{Experimental Results}
\label{experiments}

\subsection{Data Acquisition}
\label{dataacquisition}

We use the DigestPath\cite{digestpath} colonoscopy tissue segment dataset to assess the performance of our algorithm. The dataset is collected from the DigestPath2019 challenge\footnote{https://digestpath2019.grand-challenge.org/}. It contains 660 very large tissue images with an average size of $5000 \times 5000$ pixels. Each image is associated with pixel-level annotation for glandular regions. This dataset originally contained annotations for malignant lesions (250 images) only. In order to obtain a tissue segmentation mask for benign glands, we used a semi-automatic approach. For this purpose, we first trained a gland segmentation model named Mild-Net\cite{graham2019mild} on the GlaS dataset\cite{glas1,glas2}, and obtained gland segmentation masks for images with normal grades, in the DigestPath dataset which were manually refined. From these image, we extracted 1733 patches of size $512\times512$ that were later resized to $256\times256$. Out of these, we kept around 1300 patches for training (train set) and the rest for testing purpose (test set). 

The obtained masks can be binary (separating glandular region and rest) which is utilized in the proposed framework or can be ternary (glandular regions, stromal regions and background) which are used for generating synthetic masks as described in Section \ref{masksynthesis}. The stromal regions in ternary masks are obtained by performing pixel-thresholding on H\&E tissue images. The procedure to acquire bounding boxes from the gland masks collected from the digestpath dataset, and also to construct them from the input location $\overrightarrow{l}$ and size $\overrightarrow{s}$ parameters (gland layout) is given in Appendix \ref{appendixc}. 

\subsection{Model Training}

To train the whole framework, we set the target tissue image size $N=256$, input noise dimensionality for the gland specific embeddings, $dim(z)=6$, latent vector size $D=32$ and generated per gland size $B=64$. For the loss function (shown in equation \ref{losseq}), we set $\lambda_1=\lambda_2=\lambda_3=100.0$ and  $\lambda_4=\lambda_5=\lambda_6=1.0$ after cross-validation tuning.

We train all models using Adam optimizer\cite{adam} with learning rate $10^{-4}$ and batch size 1 for approx 300K iterations. For each iteration, we first update the generator weights $\{\theta_A, \theta_I, \theta_M, \theta_E\}$, then update discriminators $\{D_T, D_Z, D_G\}$. The framework is implemented in Pytorch on an Nvidia Titan X and took almost 2 days for training.

\subsection{Visual Results}

\begin{figure}[h]
\centering
\includegraphics[height=180pt, width=350pt]{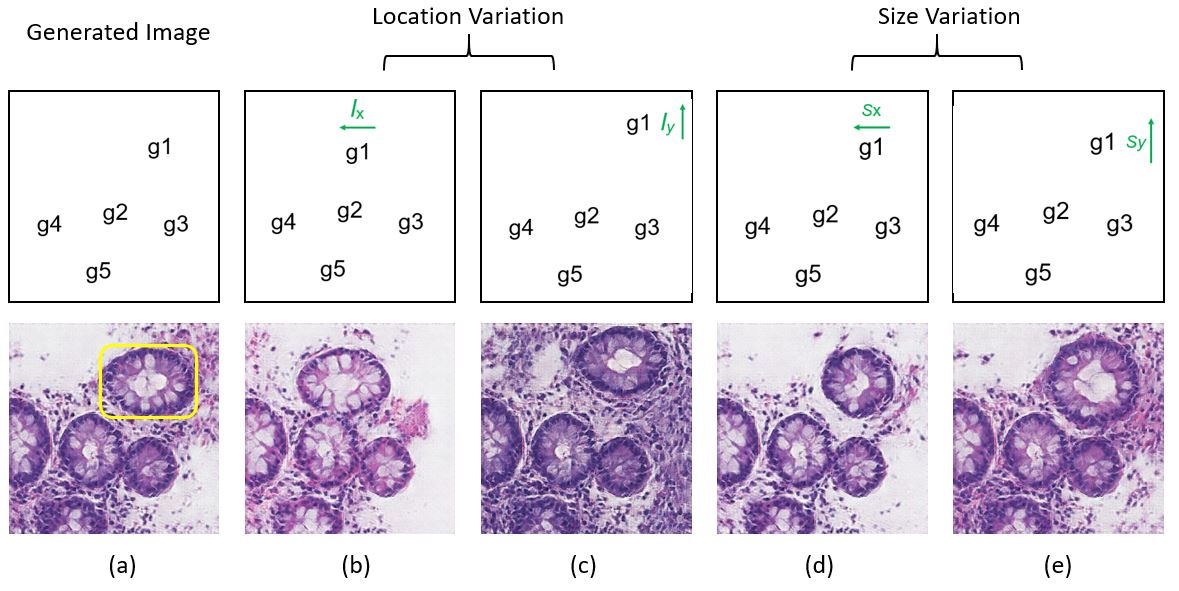}
\caption{The leftmost image (a) shows the generated sample out from the proposed framework. Images on the right to it shows the change in appearance of the yellow bordered gland, after altering location $\protect\overrightarrow{l}=(l_x,l_y)$ and size $\protect\overrightarrow{s}=(s_x,s_y)$. (b) and (c) shows the shift of that gland to left side (lowering $l_x$) and upwards (increasing $l_y$), respectively. For the same gland, (d) shows the contraction horizontally and (e) shows expansion vertically after reducing ($s_x$) and increasing ($s_y$), respectively.}
\label{usercontrol}
\end{figure}

The visual results of generated tissue images (from the test set), can be seen in Figure \ref{benign_results}. We can observe that glandular shape are preserved, tissue components like goblet cells, stromal regions are constructed with fidelity with moderate deformities in the glandular lumen. The generated tissue component masks also appear close to actual masks. The slight deformities and variations in shapes of those masks is a result of using Gaussian noise in the representation embedding of glands, which also make them realistic in nature. 
We also investigate the change in appearance of glands after altering size $\overrightarrow{s}$ and location $\overrightarrow{l}$. Figure \ref{usercontrol} shows the results of tissue images after changing their locations and sizes. The bounding boxes get modified after altering sizes and locations, which effectively change the size and orientation of glands.

\vspace{-2mm}

\subsection{Quantitative Analysis}

We evaluate the quality of generated images (from the test set) using the Frechet Inception Distance (FID)\cite{fid}, a standard metric used to assess the quality of images by the generative model. It computes the distance between convolution feature maps calculated for real and generated images. For our experiments, to collect convolution features, we use the pretrained InceptionV3\cite{inceptionv3} network trained on the ImageNet dataset\cite{imagenet}. The lower the FID score is the better is the image quality. As the metric depends on the image size, to get a sense of its scale, we also compute the FID score between ground truth images and random noise of the same size. As a baseline, we adapt the image-to-image Pix2Pix network\cite{pix2pix} to generate tissue images from existing tissue component masks, and compute the FID for tissue images generated by it. The comparative results are shown in Table \ref{fid_comparison}. 
Table \ref{fid_comparison} shows the proposed model achieves better results compared to the random noise and little inferior to that of that of Pix2Pix. The reason can be that as, we are aiming to construct gland segmentation masks as well along with the final tissue images, while Pix2Pix assumes ground truth masks already present and constructs the tissue image from it. Thus, the construction error in generation of masks can influence the performance of our framework, and may slightly lower the quality of generated images. However, looking at the scale of FID values, the difference between both frameworks is not significant. 

\begin{table}[hbt!]
\centering
\begin{tabular}{|l|l|}
\hline
Model   & FID \\ \hline
Random  & 485 ± 5.7\\ \hline
Pix2Pix & 120 ± 2.7\\ \hline
Proposed Framework    & 134 ± 2.4\\ \hline
\end{tabular}
\caption{Frechet Inception Distance (FID) score comparison}
\label{fid_comparison}
\end{table}

\begin{figure}[t]
\centering
\includegraphics[width=250pt, height=170pt]{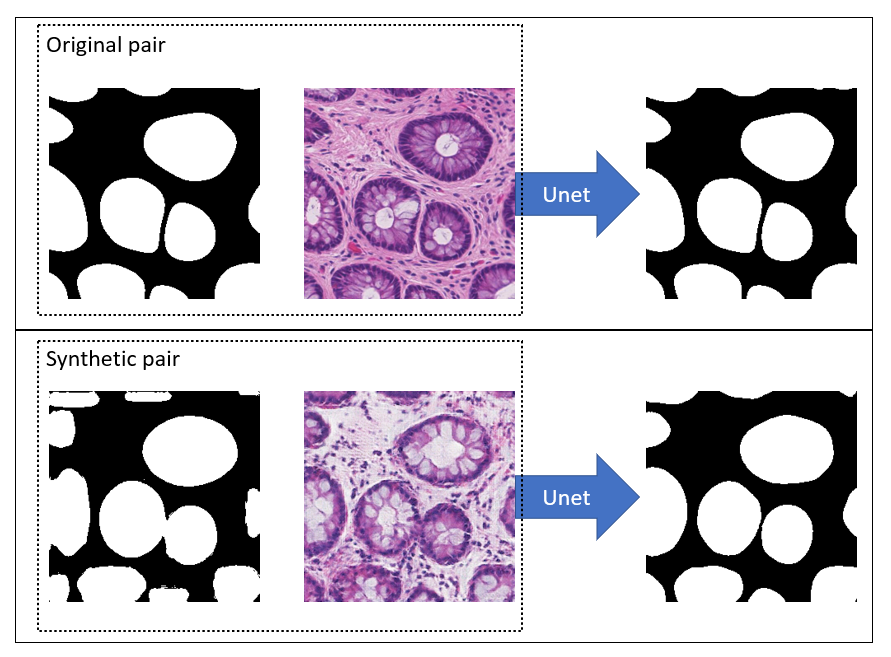}
\caption{Samples of both real (above) and constructed (below) annotated pairs of tissue images and corresponding gland segmentation masks. The masks shown on the right side are generated from the U-net based segmentation algorithm when applied on original (above) and synthetic (below) images.}
\label{unet_segmentation}
\end{figure}

\subsection{Assessment through gland segmentation}

We also assess the quality of annotated pairs generated by our framework using the U-net based gland segmentation algorithm\cite{unet}. We train U-net on patches of size $256\times256$ from the train set and compute segmentation masks of both real and synthetic images from the test set. We use the Dice score\cite{dice} between the real component masks and masks computed by U-net on real images, and also between the generated component masks and masks computed by U-net on synthetic images. Sample results are shown in Figure \ref{unet_segmentation}.

We obtained an average Dice index of 0.9022 (with standard deviation 0.006) and 0.9001 (with standard deviation 0.012) for both respective cases. This highly similar obtained score validates the applicability of both generated tissue images along with their tissue component masks, for the evaluation of gland segmentation algorithms.

\subsection{Image Synthesis from Masks Generated Using Latent Diffusion Model}

\begin{figure}[hbt!]
\centering
\includegraphics[width=380pt, height=180pt]{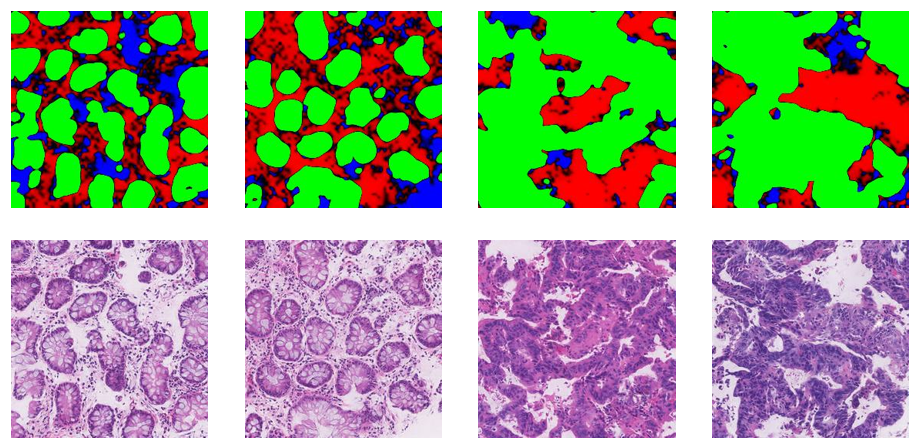}
\caption{Annotated pairs synthesized using the latent diffusion model.  The first two pairs (from right) depict annotated colon tissue images of the benign grade, while the last two show images of the malignant type.} 
\label{diffusion_images_annotated}
\end{figure}

The tissue component masks generated through the latent diffusion model based framework (described in Section \ref{masksynthesis}), along with the corresponding tissue images created using the residual encoder-decoder network\cite{interactionscenegraph}, are displayed in Figure \ref{diffusion_images_annotated}. The resulting FID between genuine images and images produced through this method is 130.5 (±2.8). Comparing these FID values with those obtained previously, it's evident that the synthetic images closely resemble real images.

\section{Conclusion and Future Directions}

In this work, we presented an interactive framework to generate annotated pairs of colorectal tissue images along with their tissue component masks. We performed experiments on DigestPath dataset and demonstrate the framework's ability to generate realistic images preserving morphological features including stroma, goblet cells and glandular lumen. The generated images maintain good FID scores when compared with the state-of-the-art image-to-image translation model. We showed variability in glandular appearance after altering sizes and locations of glands. Additionally, we also demonstrated the applicability of synthetic annotated pairs for the evaluation of gland segmentation algorithms.

One visible limitation in the proposed framework is that it requires glandular layout to construct images which still need some efforts. However, we have given an alternate methodology to construct the glandular masks and thereby generating tissue images.

The idea can be extended in future to generate annotated tissue image pairs for various tasks in computational histopathology such as nuclei segmentation, cancer grading etc. The generated pairs can potentially replace the real-world training data, pass the legal and security barriers while using them, assist the training and validation of digital pathology algorithms, and reduce the cost and efforts of acquiring data.

\newpage

\bibliographystyle{spiebib}
\bibliography{sample}

\newpage 

\section*{Appendix}
\label{appendix}

Here we describe the neural network architectures for each component within the proposed framework

\subsection{Individual Mask Generator Network}

We generate the individual mask for each of the glands using the \textit{individual mask generator network}. The input is the individual gland latent vectors obtained after affine transformation on the original gland embeddings, and output is the $64 \times 64$ pixels glandular mask with all elements ranged between 0 and 1. The mask regression network composed of series of individual mask generator blocks, where each block consist of interpolation + convolution + batch normalization + ReLU activation operations. The exact architecture is shown in the table \ref{tabletwo}, while architecture of the \textit{individual mask generator block} is shown in Table \ref{three}.

\begin{table}[hbt!]
\centering
\begin{tabular}{|c|c|c|c|}
\hline
\textbf{Index} & \textbf{Inputs} & \textbf{Operation}                            & \textbf{Output Shape} \\ \hline
(1)            & -               & Gland Latent Vector                        & 32                    \\ \hline
(2)            & (1)             & Reshape                                       & 32 x 1 x 1            \\ \hline
(3)            & (2)             & Individual Mask Generator Block                          & 32 x 2 x 2            \\ \hline
(4)            & (3)             & Individual Mask Generator Block                          & 32 x 4 x 4            \\ \hline
(5)            & (4)             & Individual Mask Generator Block                          & 32 x 8 x 8            \\ \hline
(6)            & (5)             & Individual Mask Generator Block                          & 32 x 16 x 16          \\ \hline
(7)            & (6)             & Individual Mask Generator Block                          & 32 x 32 x 32          \\ \hline
(8)            & (7)             & Individual Mask Generator Block                          & 32 x 64 x 64          \\ \hline
(9)            & (8)             & Conv2d (K=1, 32 $\rightarrow$ 1) & 1 x 64 x 64           \\ \hline
(10)            & (9)             &  Sigmoid & 1 x 64 x 64           \\ \hline
\end{tabular}
\caption{Architecture of the individual mask generator network. The function implements function $I$ from the main text.  The notation Conv2d(K , $C_{in}$ → $C_{out}$) is a convolution with $K \times K$ kernels, $C_{in}$ input channels and $C_{out}$ output channels; all convolutions with stride 1 with zero padding that ensures input and output have the same spatial size.}
\label{tabletwo}
\end{table}

\begin{table}[hbt!]
\centering
\begin{tabular}{|c|c|}
\hline
\textbf{Operation}                             & \textbf{Output Shape} \\ \hline
Interpolation                                  & 32 x 2S x 2S          \\ \hline
Conv2d (K=3, 32 $\rightarrow$ 32) & 32 x 2S x 2S          \\ \hline
Batch Normalization                            & 32 x 2S x 2S          \\ \hline
ReLU                                           & 32 x 2S x 2S          \\ \hline
\end{tabular}
\caption{Architecture of the individual mask generator block. The input is the feature map of shape $C \times S \times S$, where C is the number of channels from the feature map of the last layer, and $S \times S$ is the dimension of height and width.}
\label{three}
\end{table}

\subsection{Mask Generator Network}

The generated intermediate tensor (explained in \ref{sectioncomponentmask}) has 32 channels, which got reduced to 1 using the \textit{mask generator network}, forming the glandular mask or tissue component mask. The network comprised of a series of convolution + Relu operations. The exact architecture is shown in Table \ref{maskgeneratornetwork}. 

\begin{table}[hbt!]
\centering
\begin{tabular}{|c|c|c|c|}
\hline
\textbf{Index} & \textbf{Inputs} & \textbf{Operation}                             & \textbf{Output Shape} \\ \hline
(1)            & -               & Generate Cumulative Mask                        & 32 x 256 x 256        \\ \hline
(2)            & (1)             & Conv2d (K=3, 32 $\rightarrow$ 16) & 16 x 256 x 256        \\ \hline
(3)            & (2)             & LeakyReLU                                      & 16 x 256 x 256        \\ \hline
(4)            & (3)             & Conv2d (K=3, 16 $\rightarrow$ 8)  & 8 x 256 x 256         \\ \hline
(5)            & (4)             & LeakyReLU                                      & 8 x 256 x 256         \\ \hline
(6)            & (5)             & Conv2d (K=3, 8 $\rightarrow$ 4)   & 4 x 256 x 256         \\ \hline
(7)            & (6)             & LeakyReLU                                      & 4 x 256 x 256         \\ \hline
(8)            & (7)             & Conv2d (K=3, 8 $\rightarrow$ 4)   & 1 x 256 x 256         \\ \hline
(9)            & (8)             & LeakyReLU                                      & 1 x 256 x 256         \\ \hline
\end{tabular}
\caption{Architecture of the mask generator network. The network implements function $M$ from the main text. LeakyReLU uses a negative slope coefficient of 0.2}
\label{maskgeneratornetwork}
\end{table}

\subsection{Encoder-Decoder Network}

The final tissue image is generated from the generated glandular mask with the help of \textit{encoder decoder network}. The encoder consist of a series of ``Encode" blocks (shown in Table \ref{encoderblock}) and generates the lower sized encoding of the input mask, while decoder comprised of a series of ``Decode" blocks (shown in Table \ref{decodeblock}) and generates the final tissue image from the encoding. The exact architecture of the encoder-decoder network is shown in Table \ref{encoderdecodernetwork}. 

\begin{table}[hbt!]
\centering
\begin{tabular}{|c|c|c|c|}
\hline
\textbf{Index} & \textbf{Inputs} & \textbf{Operation}                             & \textbf{Output Shape} \\ \hline
(1)            & -               & Generate Component Mask                        & 1 x 256 x 256         \\ \hline
(2)            & (1)             & Encode(1,64)                                   & 64 x 128 x 128        \\ \hline
(3)            & (2)             & Encode(64,128)                                 & 128 x 64 x 64         \\ \hline
(4)            & (3)             & Encode(128,256)                                & 256 x 32 x 32         \\ \hline
(5)            & (4)             & Encode(256,512)                                & 512 x 16 x 16         \\ \hline
(6)            & (5)             & Encode(512,512)                                & 512 x 8 x 8           \\ \hline
(7)            & (6)             & Encode(512,512)                                & 512 x 4 x 4           \\ \hline
(8)            & (7)             & Encode(512,512)                                & 512 x 2 x 2           \\ \hline
(9)            & (8)             & Encode(512,512)                                & 512 x 1 x 1           \\ \hline
(10)           & (9,8)           & Decode(512,512)                                & 1024 x 2 x 2          \\ \hline
(11)           & (10,7)          & Decode(1024,512)                               & 1024 x 4 x 4          \\ \hline
(12)           & (11,6)          & Decode(1024,512)                               & 1024 x 8 x 8          \\ \hline
(13)           & (12,5)          & Decode(1024,512)                               & 1024 x 16 x 16        \\ \hline
(14)           & (12,4)          & Decode(1024,256)                               & 512 x 32 x 32         \\ \hline
(15)           & (14,3)          & Decode(512,128)                                & 256 x 64 x 64         \\ \hline
(16)           & (15,2)          & Decode(256,64)                                 & 128 x 128 x 128       \\ \hline
(17)           & (16)            & Upsample                                       & 128 x 256 x 256       \\ \hline
(18)           & (17)            & Conv2d (K=4, 128 $\rightarrow$ 3) & 3 x 256 x 256         \\ \hline
(19)           & (18)            & Tanh                                           & 3 x 256 x 256         \\ \hline
\end{tabular}
\caption{Architecture of the encoder-decoder network. The network implements the function $E$ from the main text.}
\label{encoderdecodernetwork}
\end{table}

\begin{table}[hbt!]
\centering
\begin{tabular}{|c|c|}
\hline
\textbf{Operation}                                & \textbf{Output Shape} \\ \hline
Conv2d (K=4, $C_{in} \rightarrow C_{out})$ & $C_{out}$ x S x S          \\ \hline
Instance Normalization (if normalize=True)        & $C_{out}$ x S x S          \\ \hline
LeakyReLU                                         & $C_{out}$ x S x S          \\ \hline
Dropout (if dropout=True)                         & $C_{out}$ x S x S          \\ \hline
\end{tabular}
\caption{Architecture of the ``Encode" block. LeakyReLU uses a negative slope coefficient of 0.2. The input is image of size $C_{in} \times S \times S$}
\label{encoderblock}
\end{table}

\begin{table}[hbt!]
\centering
\begin{tabular}{|c|c|}
\hline
\textbf{Operation}                                        & \textbf{Output Shape} \\ \hline
ConvTranspose2d(K=4, $C_{in} \rightarrow C_{out}$) & $C_{out}$ x S x S          \\ \hline
Instance Normalization                                    & $C_{out}$ x S x S          \\ \hline
ReLU                                                      & $C_{out}$ x S x S          \\ \hline
Dropout (if dropout=True)                                 & $C_{out}$ x S x S          \\ \hline
\end{tabular}
\caption{Architecture of the ``Decode" block. The input is image of size $C_{in} \times S \times S$}
\label{decodeblock}
\end{table}

\subsection{Mask and Image Discriminators}

The discriminator we employed has the similar architecture for both glandular masks ($D_M$) and tissue images ($D_Z$), takes the real or fake image of shape $C \times 256 \times 256$ as an input ($C=1$ for component mask and $C=3$ for tissue image), and classifies an overlapping grid of size $7 \times 7$ image patches from the input image as real or fake. The exact architecture of the discriminator is shown in Table \ref{discriminatornetwork}

\begin{table}[]
\centering
\begin{tabular}{|c|c|c|c|}
\hline
\textbf{Index} & \textbf{Inputs} & \textbf{Operation}                                    & \textbf{Output Shape} \\ \hline
(1)            & -               & Generate the Image                                    & C x 256 x 256         \\ \hline
(2)            & (1)             & Conv2d (K=4, C $\rightarrow$ 16, S=2)    & 16 x 128 x 128        \\ \hline
(3)            & (2)             & LeakyReLU                                             & 16 x 128 x 128        \\ \hline
(4)            & (3)             & Conv2d (K=4, 16 $\rightarrow$ 32, S=2)   & 32 x 64 x 64          \\ \hline
(5)            & (4)             & LeakyReLU                                             & 32 x 64 x 64          \\ \hline
(6)            & (5)             & Instance Normalization                                & 32 x 64 x 64          \\ \hline
(7)            & (6)             & Conv2d (K=4, 32 $\rightarrow$ 64, S=2)   & 64 x 32 x 32          \\ \hline
(8)            & (7)             & LeakyReLU                                             & 64 x 32 x 32          \\ \hline
(9)            & (8)             & Instance Normalization                                & 64 x 32 x 32          \\ \hline
(10)           & (9)             & Conv2d (K=4, 64 $\rightarrow$ 128, S=2)  & 128 x 16 x 16         \\ \hline
(11)           & (10)            & LeakyReLU                                             & 128 x 16 x 16         \\ \hline
(12)           & (11)            & Instance Normalization                                & 128 x 16 x 16         \\ \hline
(13)           & (12)            & Conv2d (K=4, 128 $\rightarrow$ 256, S=2) & 256 x 8 x 8           \\ \hline
(14)           & (13)            & LeakyReLU                                             & 256 x 8 x 8           \\ \hline
(15)           & (14)            & Instance Normalization                                & 256 x 8 x 8           \\ \hline
(16)           & (15)            & Conv2d (K=4, 256 $\rightarrow$ 1 , S=1)  & 1 x 7 x 7             \\ \hline
\end{tabular}
\caption{Architecture of the Discriminator network. C=1 when input is the tissue component mask and C=3 for the tissue image. All but the last Conv2d operation has stride 2. LeakyReLU uses a negative slope coefficient of 0.2}
\label{discriminatornetwork}
\end{table}

\subsection{Gland Discriminator}

Our gland discriminator $D_G$ consumes image pixels corresponding to glandular areas from the real or fake tissue images, and classifies them as real or fake. The glandular areas are cropped out using their bounding box coordinates, and resized to $64 \times 64$ pixels using the bilinear interpolation method. The exact architecture of the gland discriminator is shown in Table \ref{glanddiscriminatornetwork}

\begin{table}[]
\centering
\begin{tabular}{|c|c|c|c|}
\hline
\textbf{Index} & \textbf{Inputs} & \textbf{Operation}                                  & \textbf{Output Shape} \\ \hline
(1)            & -               & Crop glandular portions from the generated image    & 3 x 64 x 64           \\ \hline
(2)            & (1)             & Conv2d (K=5, 3 $\rightarrow$ 16, S=2)  & 16 x 30 x 30          \\ \hline
(3)            & (2)             & Batch Normalization                                 & 16 x 30 x 30          \\ \hline
(4)            & (3)             & LeakyReLU                                           & 16 x 30 x 30          \\ \hline
(5)            & (4)             & Conv2d (K=5, 16 $\rightarrow$ 32, S=2) & 32 x 13 x 13          \\ \hline
(6)            & (5)             & Batch Normalization                                 & 32 x 13 x 13          \\ \hline
(7)            & (6)             & LeakyReLU                                           & 32 x 13 x 13          \\ \hline
(8)            & (7)             & Conv2d (K=5, 32 $\rightarrow$ 64, S=2) & 64 x 5 x 5            \\ \hline
(9)            & (8)             & Global Average Pooling                              & 64                    \\ \hline
(10)           & (9)             & Affine Transformation                               & 1024                  \\ \hline
(11)           & (10)            & Affine Transformation                               & 1                     \\ \hline
\end{tabular}
\caption{Architecture of the gland discriminator, $D_G$. LeakyReLU uses a negative slope coefficient of 0.2}
\label{glanddiscriminatornetwork}
\end{table}

\subsection{Acquisition of Bounding Boxes}
\label{appendixc}

After we collect the tissue images and their annotated gland masks from the digestpath dataset as described in section \ref{dataacquisition}, we used the OpenCV (Open Source Computer Vision Library) python library\cite{opencv} to extract the location of glands i.e., centroids of white blob objects from the black and white tissue component mask (as shown in mask in figure \ref{scene_generation_idea}). Later we also collected the bounding boxes of those identified glandular objects using the built-in function boundingRect() function of the same library.

During inference, apart from using the ground truth bounding boxes obtained by the procedure described above, we also construct  bounding box for the gland $g_k$ using the input size $\overrightarrow{s_k}$ and location $\overrightarrow{l_k}$ attributes, taken from the gland layout. Given the input size $\overrightarrow{l_k} = (s_{kx},s_{ky})$, where $s_x$ and $s_y$ are the horizontal and vertical spanning lengths of glands, and the centroid location $\overrightarrow{l_k}=(l_{kx},l_{ky})$, the bounding box coordinates for $g_k$ are computed as,

\begin{equation}
    b_k = (l_{kx} - (s_{kx}/2), (l_{ky} - (s_{ky}/2), (l_{kx} + (s_{kx}/2), (l_{ky} + (s_{ky}/2)).
\end{equation}

Overall, input to the proposed framework is the set of glandular locations, their sizes, their bounding boxes acquired from the dataset or constructed using the above procedure, and output is the pair of the tissue image and its tissue component mask.

\end{document}